\documentclass[11pt]{article}

\usepackage{amssymb} 
\usepackage{amscd}

\newcommand {\newsection}{\setcounter{equation}{0}\section}

\def \be {\begin{equation}}
\def \ee {\end{equation}}
\def \bea {\begin{eqnarray}}
\def \eea {\end{eqnarray}}
\def \ba {\begin{array}}
\def \ea {\end{array}}
\def \pp {{\Bbb P}}
\def \zz {{\Bbb Z}}
\def \cc {{\Bbb C}}
\def \rr {{\Bbb R}}

\def \ff {{\cal F}}
\def \EE {{\cal E}}
\def \dd {{\cal D}}
\def \oo {{\cal O}}
\def \nn {{\cal N}}

\def \CY {Calabi-Yau }

\def \SYM {super Yang-Mills }
\newcommand {\mat}[1]{\mathop{\rm #1}\nolimits}

\def \End {\mathop{\rm End}\nolimits}

\newcommand{\Hom}{\mathop{\rm Hom}\nolimits}

\begin{document}

\begin{titlepage}
\titlepage
\rightline{hep-th/0104041}
\rightline{CPHT-S012.0301}
\rightline{SISSA 25/2001/FM}
\vskip 3cm
\centerline{{ \bf \large Variations on stability}}
\vskip 1.5cm
\centerline{ Ruben Minasian$^a$ and  Alessandro Tomasiello$^b$}
\begin{center}   
\em $^a$Centre de Physique Th{\'e}orique, Ecole
Polytechnique\footnote{Unit{\'e} mixte du CNRS et de l'EP, UMR 7644}
\\91128 Palaiseau Cedex, France\\{\tt Ruben.Minasian@cpht.polytechnique.fr}  
\end{center}
\begin{center}
\em $^b$International School for Advanced Studies (SISSA)
\\via Beirut 2-4, 34014 Trieste, Italy\\{\tt tomasiel@sissa.it}
\end{center}
\vskip 1.5cm
\begin{abstract} 

We explore the effects of non-abelian dynamics of D-branes on their 
stability and introduce Hitchin-like modifications to previously-known
stability conditions. The relation to brane-antibrane systems is used in
order to rewrite the equations in terms of superconnections and
arrive at deformed vortex equations.

\end{abstract}   

\vfill
\begin{flushleft}
{\today}\\
\end{flushleft}
\end{titlepage}

\newpage 

\newsection{Introduction and Discussion}

The non-abelian nature of D-branes is one of the key 
features of their modern understanding. The connection with non-abelian
gauge theories has served as a natural basis for many explorations. In
this paper we concentrate on the effects of non-abelian dynamics of
D-branes on their BPS stability.

It is well-known that stability plays an essential role in the study of
moduli spaces of BPS states. To see this, it is already sufficient to look
at the limit in which the volume of the ambient manifold is large; in that
limit, branes are described by specifying a (non-trivially) embedded
manifold in the spacetime and a gauge connection on it, or to put shortly
supersymmetric cycle. Typically, in the study of BPS conditions, one first
gets various types of conditions on the cycles, such as being
holomorphically embedded or being a special lagrangian submanifold. In
addition to that, one gets equations which involve the curvature $F$ of
the gauge connection $A$ on the brane. These equations, which in the
holomorphic case are the Hermitian-Yang-Mills equations, do not always
admit a solution. However one typically is able to formulate a condition
for having solutions on the connection $A$ in purely algebraic-geometric
terms, involving some inequality on subbundles.  This condition is called
stability; it is nice that, in a sense, the mathematical definition has
anticipated the physical meaning.

Away from the large volume situation, this situation
will get corrections; the equations will be deformed, and in fact the whole
geometrical interpretation of branes will lose its validity. Steps towards
understanding these corrections have been done in two approaches. The first 
one is to redo the BPS analysis starting from the complete Dirac-Born-Infeld
plus Chern-Simons action \cite{Marino:2000af}, and gives a deformation 
of the Hermitian-Yang-Mills equations:
\be
[\mat{Im}( e^{i \theta} e^{F + \omega} )]_{\mat{top}} = 0,
\label{zzz}
\ee
where $\omega$ is the K{\"a}hler form, and $\theta$ is a phase.
The second, more abstract, approach, 
which avoids the difficulty of finding correct
equations, directly introduces the notion of a deformed stability
condition \cite{Douglas:2000ah}. 

We recall now that in earlier days of D-branes considerable attention was
paid to yet another type of modifications to Yang-Mills equations
involving scalars (see. e.g. \cite{Bershadsky:1996qy, Harvey:1998gc} for
context closely related to ours). These equations -- the Hitchin equations,
or rather generalized Hitchin equations as they do not have to be
considered only in two dimensions -- also lead to a certain notion of
stability. The recent progress in understanding the
non-abelian dynamics of D-branes  \cite{Myers:1999ps} warrants
revisiting stability conditions
with intention of incorporating the scalars. The resulting modification of
stability while rather mild for the whole BPS moduli space, may turn out
to be rather dramatic for certain individual cycles. In particular we
will see that some previously discarded cycles may get rehabilitated under
the modified stability conditions.

The connection between transverse scalars and tachyons (superconnections)
is natural and intuitive -- both serve the purpose of ``localization"; more
precisely, we will argue that the tachyon can be expressed in terms of 
the characteristic polynomial of the transverse scalars (see eq. (\ref{char})
below). This
connection is a key to a better conceptual understanding of the
generalizations of (\ref{zzz}).
These can be compactly summarized as
deformed vortex equations
\[
\mat{Im}\left( \, \mat{Sym} \{ \exp
\left(\ba {cc}i
\theta_1\mat{Id}_{N_1}&0\\0&i\theta_2\mat{Id}_{N_2}\ea\right)\,
[e^{[\dd,\bar\dd]}e^{\omega + J}]_{\mat{top}} \}\, \right) = 0 
\]
in terms of a sort of holomorphic ``contracting'' 
superconnection $\dd$ ($\bar\dd$), and of a real two-form $J$ introduced for
bookkeeping, as we will explain in section \ref{dte}.

The deformed equations we are studying here would seem to suggest, along the 
lines of \cite{Hitchin:1987vp,Hitchin2}, the presence of an integrable system
on the cotangent bundle to the moduli space of $\Pi$-stable branes, giving
thus a deformation to the so-called Hitchin integrable system \cite{Hitchin2};
this could have some contact with \cite{DEL}, but we will not try
address this question here.

A brief outline of the paper is the following. Section 2 is a more
technical introduction, containing a brief review of the deformed
Hermitian Yang-Mills equation and Hitchin equations. A generalization of
the former in the spirit of the latter (or vice versa) is presented in
section 3. The
interpretation of the new equations and their implications for stability
are discussed in section 4. Finally, in section 5 we explore the relation
with brane-antibrane systems and with
superconnections.

\newsection{Preliminaries}
\label{pre}
Let us  start with a simple example. Consider \SYM in $d=4$ with
$\nn=4$. In this case, one can easily verify that the condition for 
preserving
half of supersymmetry, if the transverse scalars $X$ are set to their
vacuum values, is
the instanton equation $F = * F$. In complex coordinates $z_1, z_2$, this reads
\be
F_{1\bar1} + F_{2\bar2} = c\,\mat{Id}, \qquad F_{12}=0
\label{sd4}
\ee
where 
$F_{ij}\equiv F_{z_i, z_j}$, $F_{i\bar j}\equiv F_{z_i, \bar z_j}$, and
another equation $F_{\bar 1 \bar2}=0$ follows from $A$ being antihermitian.
Let us now consider the T-duals of these equations. 
The general procedure is the same as that of dimensional reduction, and reads $D_i \to
X_i$:  a connection becomes an endomorphism (a matrix).
More precisely, this rule, for T-duality, means that we are but rewriting
the covariant derivatives as infinite dimensional matrices; then, the 
expression obtained in this way will be true also in the case in which
matrices are finite dimensional. As mentioned, the outcome of this
reasoning is the same
as that of dimensional reduction. If T-duality is in the
$z_2, \bar z_2$ directions, the result is 
\be
F_{1\bar1} + [X, X^\dagger] = c \,\mat{Id}, \qquad D_1 X=0
\label{sdH}
\ee
where we have called $X_2\equiv X$. These equations can be obtained also by 
considering \SYM in $d=2$ and with $\nn=(8,8)$, which is indeed a dimensional 
reduction, or T-dual, of $d=4$, $\nn=4$, and looking for solutions which 
preserve half supersymmetry with one complex scalar turned on. For these
reasons they 
have already been argued to be relevant in several physical situations 
\cite{Bershadsky:1996qy,Bershadsky:1995vm,Vafa:1994tf,Giddings,mst,Bonelli:1999it}; on 
the mathematical side, they have been studied by Hitchin 
\cite{Hitchin:1987vp}.

One can look for more general BPS conditions in \SYM theories starting instead 
from the $d=10$, $\nn=1$ case, keeping all the complex scalars on:
\be
F_{1\bar1} + \ldots + F_{5\bar5} = c\,\mat{Id}, \qquad F_{ij}=0.
\label{10d}
\ee
Again, reducing these to lower dimensions gives equations involving
complex scalars in the adjoint: for instance in 4 dimensions one
gets
\bea
\nonumber &&F_{1\bar1} + F_{2\bar2} + 
[X_1 , X_1^\dagger] + [ X_2 , X_2^\dagger] + [X_3 , X_3^\dagger]
= c\,\mat{Id}; \qquad F_{12}=0;\\
 &&D_i X = 0, \ \ i=1,2; \qquad [X_a, X_b]=0, 
\ \ a,b= 1,2,3.
\label{sdH4}
\eea
These are a modification of the usual self-duality conditions (\ref{sd4})
in 4 dimensions.

These \SYM theories are known to describe the dynamics of flat branes in 
$\rr^{10}$ in zero slope limit. One can wonder how does the situation change 
away from this limit. Let us first start from the abelian case. There, one 
knows that the effective action becomes the Dirac-Born-Infeld; putting this
together with the Chern-Simons term, describing coupling to RR fields, one 
obtains an action whose BPS conditions can be studied 
\cite{Marino:2000af}. To describe the 
result, let us first covariantize the equations we have obtained so far in the 
case {\sl without scalars}. All of them can be written in the form 
\be
F\wedge \frac{\omega^{n-1}}{(n-1)!} = c \,
\frac{\omega^n}{n!} \, \mat{Id}, \qquad F^{(2,0)}=0,
\label{hym}
\ee
where $n=d/2$ is the complex dimension of the brane we are considering,
and $c$ is a constant. In this
form, they are known as Hermitian-Yang-Mills (HYM), or Donaldson-Uhlenbeck-Yau 
\cite{uhlyau}, 
equations. The BPS conditions gotten from DBI + CS can be instead written in
the compact form 
\be 
[\mat{Im}( e^{i \theta} e^{ F + \omega} )]_{\mat{top}} = 0, \qquad
F^{(2,0)}=0.
\label{abmmms}
\ee
The subscript $_{\mat{top}}$ indicates that we only  have to take the
top-form part 
of the expansion $e^{F + \omega}$. As one can see the second equation, 
which is the holomorphicity condition, is not changed. This agrees
with the decoupling conjecture of \cite{Brunner:2000jq} stating that
for B-type branes the superpotential does not depend on K{\"a}hler moduli.
The F-flatness condition $F^{(2,0)}=0$ is almost trivial to reduce, and
often we will not write
it or its lower-dimensional descendants explicitly.

For instance, in 4 dimensions, first equation in (\ref{abmmms}) 
can be written as
\be
i F\wedge \omega = 
\mat{tg}(\theta) \left( \frac12 F^2 + \frac12 \omega^2 \right); 
\label{4dmmms}
\ee
here we have used the fact that $F$ is
anti-hermitian and so, in the abelian case, purely imaginary, and that 
$\omega$ is real. The $(1,1)$ part of the equations (\ref{hym}) 
in 4 dimensions (the self duality 
equations) can be seen as a linearized in $F$
version of (\ref{4dmmms}), with $c=-i/2 \mat{tg}\theta$.

Strictly speaking, the equations were derived only in the abelian case.
In the non-abelian case, the action is not completely known.
Nevertheless, (\ref{abmmms}) admit a very natural non-abelianization
by treating $F$ as a matrix, and putting an identity in the $\omega^n$
part. This will clearly be well-defined, thanks to the transformation of the
curvature under gauge transformation $F \to U F U^\dagger$.
In fact, we can write this non-abelianized version again in a compact
form as 
\be
[\mat{Im}( e^{i \theta} e^{F + \omega \mat{Id}} )]_{\mat{top}} = 0;
\label{mmms}
\ee
this has now to be read with a little caveat, namely that $\mat{Im}(\cdot)$ 
is now $( \, (\cdot) - (\cdot)^\dagger \, )/ 2i$ (this is because now $F$ 
is not purely imaginary, but anti-hermitian). Practically, this means that 
for instance in 4 dimensions we get again the equations (\ref{4dmmms}).
Notice that the equations come
automatically with a symmetrizer, due to the fact that they are written in
terms of forms. 

So far we have seen that:
\begin{itemize}
\item[-] the BPS conditions in the zero slope limit, in which there is a 
\SYM description, are given by dimensional reductions of the 10 dimensional 
equations (\ref{10d}), which can be rewritten as $F\omega^4 = 
c \omega^5$ (from 
now on we will omit $\wedge$ and Id if there is no danger of confusion).
\item[-] away from the zero slope limit, the deformation to these equations are
known in all dimensions, and in particular in 10 dimensions.
\end{itemize}
It is natural, then, to assume that dimensional reduction of the deformed 
equations (\ref{mmms}) gives the correct BPS conditions in lower
dimensions with transverse complex scalars turned on. This will give a
deformation of the  Hitchin-like equations we have shown before to arise
from super  Yang-Mills, for instance 
(\ref{sdH}) and (\ref{sdH4}). 

Before we go on and find the general deformed equations with scalars, let us 
write also first equations in (\ref{sdH}) and (\ref{sdH4}) 
in a covariant  form. For (\ref{sdH})
it is easier: a simple way is to write it is
\be
F -i [X, X^\dagger]\omega= c\,\omega.
\label{Hcov}
\ee
We could have as well made $X$ a one form; this second choice 
is the version studied by Hitchin, and we will have more to say about this in
section \ref{twi}.
As to the second, (\ref{sdH4}), 
there are more scalars in this case.
Let us then 
introduce a transverse form $\omega_\perp$, with which the indices of the 
$X_a$ and of the $X_{\bar a}\equiv X^\dagger$, 
which are scalars in the spacetime but vectors in the transverse 
directions, can contract. 
This is a very natural thing to do, in light of the 10 dimensional origin
of the equations. 
Then we can write (\ref{sdH4}) in the form
\be
F \omega  + \frac{\omega^2}2 \, (i_X + i_{X^\dagger})^2 \omega_\perp = 
c \frac{\omega^2}2,
\label{Hcov4}
\ee
where $i_X \equiv X^a i_{e_a}, i_{X^\dagger}\equiv X^{\bar a} i_{e_{\bar a}}$ 
are contractions with the holomorphic and
antiholomorphic parts $X, X^\dagger$ of the transverse scalars. Notice that
$i_X^2, i_{X^\dagger}^2$ are zero because they reproduce the commutators
$[X_a,X_b]$ and $[X_{\bar a}, X_{\bar b}]$; $i_X i_{X^\dagger}+
i_{X^\dagger}i_X$
then gives the desired combination of commutators. Notice that, in order to 
keep $\omega_\perp$ real, we have chosen it to be of the form 
$i \sum_a dx^a d\bar x^a$. 

We confine ourselves to a holomorphic setup, and thus always consider
the case when both the original equations and the reductions are
even-dimensional.  In principle we could have as well
reduced an odd number of covariant derivatives. Even though we will not do
this here let us notice that such reductions could also lead to
interesting equations. For instance, the odd-codimension reductions
of self-duality equations yields two important equations: Nahm equations
in 0+1 dimension, and the Bogomol'nyi equations for monopoles in 2+1
dimensions.  Another interesting situation involving an odd-dimensional
reduction is the gauge theory on the coassociative cycles in manifolds of
$G_2$ holonomy.

\newsection{The equations}
\label{eqns}

 Let us now turn to the reduction of the equations (\ref{mmms}). For
illustrative purposes, we will first do the reduction from 4 dimensions to
2, which can be viewed as reducing from 10 to 2 but with only one complex
scalar on.  Then we will tackle the case with all the complex scalars on
and reduce from 10 dimensions to 4. After that we will cast the result in
a dimension-independent form analogous to that of (\ref{mmms}).
The most convenient four-dimensional starting point  is in the form 
(\ref{4dmmms}). 
Reducing this in 2 dimensions one gets
\be
\frac12 \{F (-i) [X, X^\dagger ] -i[X, X^\dagger] F 
-i(DX^\dagger \bar DX - \bar DX DX^\dagger)\}
+ \omega = 
i \mat{tg}(\theta) (F -i [ X, X^\dagger ] \omega);
\label{refsym}
\ee
this is, as expected, a deformation (by the first three 
terms) of (\ref{Hcov}): these two equations are the same in linear order.

Coming now to the more laborious task of reducing from 10 to 4, we 
will not actually perform the  reduction of the equations, but
reduce instead the expression
\[
[e^{F + \omega \mat{Id}} ]_{\mat{top}}= \frac1{5!} F^5 + 
\frac1{4!} F^4 \omega + 
\frac1{3!2!} F^3 \omega^2 +\frac1{2!3!} F^2 \omega^3 +
\frac1{4!} F \omega^4 +\frac1{5!} \omega^5 
\]
and then restore the phase $\theta$ at the end.
To write down the result in 4 dimensions, we introduce some little more 
piece of notation. Let $d_A \equiv  D + \bar D \equiv dz^i D_i + 
d\bar z^i D_{\bar i}$, where $D_i, D_{\bar i}$ are the covariant derivatives.
With the help of this we will write the result in a form halfway from an 
explicit and a contracted one, and then explain how to go on in either 
direction:

\bea
\nonumber&&\mat{Sym}\{\ \left[\left(\frac{F^2}{2!} 
\frac{ (i_X + i_{X^\dagger})^6}{3!} +
F \frac{(-)[ d_A, i_X + i_{X^\dagger}]^2}{2!}
\frac{ (i_X + i_{X^\dagger})^4}{2!} +
\frac{(-)^2[ d_A, i_X + i_{X^\dagger}]^4}{4!}(i_X + i_{X^\dagger})^2
\right)\frac{\omega_\perp^3}{3!}\right]+\\
\nonumber&&\left[\left(F \omega \frac{ (i_X + i_{X^\dagger})^6}{3!} + 
\omega \frac{(-)[ d_A, i_X + i_{X^\dagger}]^2}{2!}
\frac{ (i_X + i_{X^\dagger})^4}{2!} \right) \frac{\omega_\perp^3}{3!} 
\right.+\\
\nonumber&&
\left.
\left(\frac{F^2}{2!} \frac{ (i_X + i_{X^\dagger})^4}{2!} + 
F \frac{(-)[ d_A, i_X + i_{X^\dagger}]^2}{2!}
(i_X + i_{X^\dagger})^2 +
\frac{(-)^2[ d_A, i_X + i_{X^\dagger}]^4}{4!}
\right) \frac{\omega_\perp^2}{2!} \right]+\\
\nonumber&&
\left[ 
\frac{\omega^2}{2!}
\frac{(i_X + i_{X^\dagger})^6}{3!}\frac{\omega_\perp^3}{3!} +
\left( F \omega \frac{ (i_X + i_{X^\dagger})^4}{2!} +
\frac{(-)[ d_A, i_X + i_{X^\dagger}]^2}{2!} (i_X + i_{X^\dagger})^2 \right) 
\frac{\omega_\perp^2}{2!} +\right. \\
&&\left.\hspace{1cm}\label{long}
\left( \frac{F^2}{2!} (i_X + i_{X^\dagger})^2 +
\frac{ F (-)[ d_A, i_X + i_{X^\dagger}]^2}{2!} \right) 
\omega_\perp \right]+ \\
\nonumber&&\left[ 
\frac{\omega^2}{2!}
\frac{(i_X + i_{X^\dagger})^4}{2!}\frac{\omega_\perp^2}{2!} +
\omega \left( F  (i_X + i_{X^\dagger})^2 +
\frac{(-)[ d_A, i_X + i_{X^\dagger}]^2}{2!} \right) 
\omega_\perp  + \frac{F^2}{2!} \right] +\\
\nonumber&&\left[\frac{\omega^2}{2!}(i_X + i_{X^\dagger})^2 \omega_\perp+
F \omega \right] + \frac{\omega^2}{2!} \ \}.
\eea
A few explanations are in order. First of all, the symmetrizer $\mat{Sym}$
comes because reduction of terms like $F^n$ can give rise to terms like 
$(i_X + i_{X^\dagger})^2F^n$ or $F(i_X + i_{X^\dagger})^2F^{n-1}$ and so on 
(one can see this for instance already in (\ref{refsym})); 
instead of writing all these possibilities, we have put a symmetrizer in the
gauge part, which is the same (however, each term  $(i_X + i_{X\dagger})^2$ 
should be considered here as a unique piece; there are no odd powers 
of $i_X + i_{X\dagger}$).
As for expressions like  
$\frac{ (i_X + i_{X^\dagger})^6}{3!} \frac{\omega_\perp^3}{3!}$, we have 
exploited here 
the notation introduced at the end of previous section: each power
$(i_X + i_{X\dagger})^2$ gives commutators $-i[X_a, X_{\bar d}]$, 
and so we have
$1/(3!) \, 
\epsilon^{abc}\epsilon^{\bar d\, \bar e\, \bar f}
[X_a, X_{\bar d}][X_b, X_{\bar e}][X_c, X_{\bar f}]=
[X_1, X_1^\dagger][X_2, X_2^\dagger] [X_3, X_3^\dagger] $
$+[X_1, X_1^\dagger][X_2, X_3^\dagger] [X_3, X_2^\dagger]  + \ldots$
apart for an overall $(-i)^3$ factor.
It can be a little harder to understand terms containing 
$[ d_A, i_X + i_{X^\dagger}]$. For instance, we have 
\bea
\nonumber&&\frac{(-)^2[ d_A, i_X + i_{X^\dagger}]^4}{4!}
(i_X + i_{X^\dagger})^2 \frac{\omega_\perp^3}{3!} =\\
\nonumber && dz^1d\bar z^1\, dz^2d\bar z^2
(-)[ D_1, i_{X^\dagger}][ D_{\bar1}, i_X ]
(-)[ D_2, i_{X^\dagger}][ D_{\bar2}, i_X ]
(i_X + i_{X^\dagger})^2 \frac{\omega_\perp^3}{3!} =\\
&&\nonumber dz^1d\bar z^1 dz^2d\bar z^2
(-i)[ D_1, X_{\bar d} ] [D_{\bar 1}, X_a ]
(-i)[ D_2, X_{\bar e} ] [D_{\bar 2}, X_b ] (-i)[ X_c, X_{\bar f} ] 
\epsilon^{abc}\epsilon^{\bar d\, \bar e\, \bar f} = \\
&&\nonumber dz^1d\bar z^1 dz^2d\bar z^2
\left((-i)D_1 X_1^\dagger D_{\bar1} X_1 \,(-i) D_2 X_2^\dagger D_{\bar2} X_2 
(-i)[X_3, X_3^\dagger] + \ldots\right) .
\eea

Let us note that (\ref{long}) reduces, in the case without scalars, to $
[e^{F + \omega \mat{Id}} ]_{\mat{top}}= 1/2!\, F^2 + F \omega + 1/2!\, 
\omega^2$, 
as it should. In fact the whole formula (\ref{long}) can be rewritten in a 
nicer and more suggestive form: 
\be
 [e^{ F + i [D + \bar D, i_X + i_{X^\dagger}] + (i_X + i_{X^\dagger})^2  }
\,e^{\omega + \omega_\perp }]_{\mat{top}} \buildrel{\mat{or,\ alternatively,}}
\over=
 [e^{ F + [D , i_{X^\dagger}] + [i_X,\bar D] + (i_X + i_{X^\dagger})^2  }
\,e^{\omega + \omega_\perp }]_{\mat{top}}.
\label{nice}
\ee
At first sight, this formula may appear  a little 
strange since a non-homogeneous object is exponentiated. 
Note that formally this expression coincides with one obtained by
reduction of $F$ directly in the exponent of (\ref{mmms}). 
However, the present computation, apart from clarifying the 
meaning of such a formal expression, constitutes  a non-trivial check
since in the process of reduction, some forms become scalars. For
instance,  from a similar formal argument (reduction of the covariant
derivative) one can write the exponent of (\ref{nice}) 
in an even more compact expression, turning its two forms 
into respectively
\be
 [e^{ \frac12
[D + \bar D -i(i_X + i_{X^\dagger}), D+\bar D +i(i_X + i_{X^\dagger})]}
\,e^{\omega + \omega_\perp }]_{\mat{top}} \ \ = \ \
 [e^{ [D + i_X, \bar D + i_{X^\dagger}] }
\,e^{\omega + \omega_\perp }]_{\mat{top}}
\label{nice2}
\ee
if one understands the commutator in a super-sense: 
$i_X$ is treated as ``fermionic'' and
the usual super-Lie bracket, which is an anticommutator on two fermions,
is used.
(Moreover, one should not forget the extra signs coming from the usual 
grading of forms: for instance $[A_1, A_2] = A_1 A_2+ A_2A_1$.)
This reminds one very much of the formalism and of the
expressions appearing in the computation of
Chern characters with superconnections \cite{quillen}, 
as already noted in \cite{Periwal:2000nn} in the context of the
modified  D-brane Chern-Simons couplings \cite{Myers:1999ps}. Actually,
at this point we just observe a similarity; tachyons and $X$ scalars, 
although related by tachyon condensation, 
are not quite the same object, of course. 
We will, nevertheless, come back to 
this later, arguing for the proper place of these similarities.

We can now conclude by putting back $\theta$, and writing the equations 
in the form
\be
\mat{Im}\left( e^{i \theta} \, 
 [e^{ F + [D , i_{X^\dagger}] + [i_X,\bar D]+ (i_X + i_{X^\dagger})^2  }
e^{\omega + \omega_\perp}]_{\mat{top}} \right)=0. 
\label{niceeq}
\ee
Note that formally an expansion of (\ref{niceeq}) in $n$ dimensions
contains terms of degree $n = n_{\parallel}+ n_\perp$ in self-explanatory
notation.  Here we keep only the purely-longitudinal forms
$n=n_{\parallel}$ ($n_\perp =0$). The need of defining extra rules and
the presence of $\omega_\perp$ in
the equations is not particularly nice, however we find this form to be
the most convenient for the analysis of the next section. In
section 5 we will reinterpret this equation and write it in a more
conceptual form\footnote{We may also recall that the equations
(\ref{abmmms}) were related in \cite{Marino:2000af} to noncommutative
Hermitian Yang-Mills equations via Seiberg-Witten map. The latter
strictly speaking applies to abelian fields only. With a progress in
finding a map applicable to more general situations it would be interesting
so see if (\ref{niceeq}) can lead to a simple form of noncommutative
Hitchin  equations.}.

\newsection{Geometrical considerations}
We have, so far, dealt with flat space and branes; but we have tried to write
our formulas in the way most independent from this situation. So we can try
to extrapolate the equations to more interesting geometrical cases; let us
think of an ambient manifold $M$ which is a factor of the space in which 
string theory lives (we will mostly have in mind a \CY threefold); the rest 
will be flat Minkowski space. 
What we know is that, when one wraps several branes on a submanifold $B$ of 
the given ambient manifold $M$, the scalars 
defining transverse fluctuations of the
brane -- which, in the abelian case, would be sections of the normal 
bundle $N(B,M)$ -- become also matrices; that
is, they are now sections of $N(B,M)\otimes \End(E)$, where $E$ is the gauge 
bundle\footnote{In principle the branes can be extended or not extended in the
extra flat directions; in the latter case we can have also scalars describing
fluctuations in these directions, but we will ignore these issues altogether
in this section.}. 
To derive the equations for this more general case, one has to start from 
the analogue of (\ref{mmms}) for a brane wrapping the whole Calabi-Yau. 
For the case of CY threefolds, this can be read in eq.(3.27)$(a)$ in
\cite{Marino:2000af}: its imaginary
part is still identical to the covariantization of its flat space counterpart,
(\ref{mmms}), that we have already seen.
Then, when the \CY is 
fibred in tori over the cycle, the same logic of T-duality as in flat space
applies, and we get again our equation with scalars (\ref{niceeq}). 
Once more we have to argue as in the flat space case: T-duality gives in
principle
only a rewriting of the equations, in which $D$'s are written as infinite 
dimensional $X$'s; but then the expression obtained in this way  
is the same for finite dimensional $X$. Moreover, in this case, we
are supposing that the equations obtained in the case in which there is a
fibration in  tori will be correct even in the general case.
 The procedure is similar to that of 
\cite{Bershadsky:1996qy,Bershadsky:1995vm}.
 
In principle one may imagine another method to get the equations with
scalars, namely to
try to directly non-abelianize relevant equations in
\cite{Marino:2000af}. However,
this would be correct only if we had the complete $\kappa$-symmetric 
non-abelian action, which is not the case. So, this second method will
not
give the full equations; in particular, one cannot get terms
with commutators $[X, X^\dagger]$. 

One could wonder what the piece with $\omega_\perp$ 
appearing in (\ref{niceeq}) is supposed to
mean in this more general case. As we mentioned at the end of the section
\ref{eqns}, due to the appearance in the
equations of the combination
$\omega + \omega_\perp$, we can substitute it more
generally  directly with (the pull-back of) $\omega_M$, the K{\"a}hler form
on the ambient manifold $M$.

A point which deserves emphasizing is that now all the covariant derivatives
we have been writing in the flat space case should not
only be covariant with respect to 
the usual gauge part, but also contain a connection $a$ coming
from the  normal bundle \cite{Hassan:2000zk}. 
This is simply because, as we noted, the $X$ are now sections of
$End(E)\otimes N(B,M)$.

\subsection{Twisting}
\label{twi}
The equations we have been writing so far were all covariantized keeping 
$X$ as scalars, as required by physics. However, as we noticed after 
(\ref{Hcov}), 
mathematically 
there would have been in principle another possibility, that of making $X$ a
form. We will see here that this corresponds in fact to the possibility of 
twisting the supersymmetric brane theory. The basic idea comes from
\cite{Bershadsky:1996qy}: 
consider the case in which the ambient manifold $M$ is a K3, and
$B$ is a divisor in it. As we have said, $X$ are sections of the normal 
bundle tensor the matrix part, $N(B,M)\otimes\End(E)$; 
but in this case, due to adjunction formula, we have $N(B,M)=K_B$, 
the canonical
bundle, which is in this case nothing but the cotangent $\Omega_B$. 
So the $X$ gets
substituted in this case by a antiholomorphic one form\footnote{This is 
due to the fact that, reducing equations
like $F_{ij}=0$, we obtain actually {\sl anti}holomorphic scalars $D_i X_a$; 
see 
for instance (\ref{sdH4}).} 
 $\phi$ (that is, a $(0,1)$ form annihilated by the 
holomorphic covariant derivative: $D \phi =0$) with values in 
$\End(E)$, and the equation (\ref{sdH}) can be written in the Hitchin form
$F + [ \phi, \phi^\dagger ] = c \omega $. 

Now, it is clear that we can do the same trick when $M_{n+1}$ is a higher 
dimensional \CY and $B_n$ is again a divisor in it.
In that case, the only change is that the 
canonical bundle is now not be the cotangent: $\phi$ is a top
antiholomorphic form on $B$. 
In this way, for instance, 
equation (\ref{sdH4}) with one scalar can be covariantized in the form 
\[
F \omega  + i[ \phi, \phi^\dagger ] = c \frac{\omega^2}2.
\]

It is now very natural to wonder whether this twisting can be applied also to 
our deformed equations (\ref{nice}). Let us stick to the cases that we have
been considering so far, in which $B$ is a divisor
in a \CY $n+1$-fold $M$. In this class of cases, there is only one
complex scalar on. 
Thus, we can consider the easier equations obtained reducing 
from $n+1$ dimensions to $n$. These equations can again be summarized by 
(\ref{nice}), but now remembering that the transverse space has dimension 2. 

First of all, for terms like $[ X, X^\dagger ] (e^{F + \omega})_{\mat{top}}$ 
it is rather easy to see how we can manage the twisting. Since $X$ is now 
replaced\footnote{If the divisor were a \CY itself, this would be not really
a replacing, but rather a rearranging of the equations using a holomorphic 
Hodge dual \cite{dontho}.} 
by the two form $\phi$, the commutator is now $[\phi, 
\phi^\dagger]$, which is a top form. But there is really no great difference
between a scalar and a top form, thanks to Hodge duality; 
so we can simply, for example, contract this top form with the rest and get 
a scalar equation $[\phi, \phi^\dagger] \cdot (e^{F + \omega})_{\mat{top}}$.
It is harder to understand the twisting of terms involving $\bar DX$. 
The guiding
principle in doing this is again Hodge duality, and a local (anti)holomorphic 
Hodge duality (see previous footnote). The result is that now, wherever $\bar D
X$ appeared, now we have to use the $(0, n-1)$ form $\bar D^\dagger \phi$, 
where $\bar D^\dagger$ is the adjoint of the $\bar D$ operator.
In this way, we have that $\bar D^\dagger \phi D^\dagger \phi^\dagger$ is a 
$(n-1,n-1)$, which can be contracted with the rest again to give a scalar 
equation. The whole formula becomes thus 
\bea
\nonumber&&
\mat{Sym}\{(e^{F + \omega})_{2n} -i [ X, X^\dagger ] (e^{F + \omega})_{2n} + 
\frac i2
\left(\bar D X D X^\dagger - D X^\dagger \bar D X \right) 
(e^{F + \omega})_{2n-2}  \}\\ \nonumber &&\longrightarrow 
\mat{Sym}\{
*(e^{F + \omega})_{2n} 
+ i^{n-1} [\phi, \phi^\dagger] \cdot (e^{F + \omega})_{2n}
+ \frac{i^{n-1}}2\left(\bar D^\dagger \phi D^\dagger \phi^\dagger 
+ (-)^n D^\dagger \phi^\dagger \bar D^\dagger \phi \right) 
\cdot (e^{F + \omega})_{2n-2} \} .
\eea

\subsection{Stability}
The existence of solutions to the type of equations we have been considering
so far is usually equivalent to a mathematical concept of stability. The 
precise definition of this stability depends on the equation which one 
considers, but roughly it is always an inequality involving subbundles of the 
bundle on which the connection is defined. It is very easy to understand at 
least why such a condition is necessary. Let us start from the simple equation
$F_A = c \,\omega$ in 2 dimensions (along with $F^{(2,0)}$, as usual)
where we have explicitly indicated that $F_A$ is the 
curvature of a connection $A$ on a bundle $E$. Taking trace and
integrating,
we get $c= (2\pi i) c_1(E)/(rk(E) Vol) \equiv (2\pi i)\mu(E)/Vol$. 
To avoid use of induction, 
let us consider the case in which $rk(E)=2$. Suppose now there is a
holomorphic subbundle (in this case it can only be a line bundle) 
$L \buildrel s\over\hookrightarrow E$, with a 
connection $A'$ on it (on a line bundle we can choose it to have constant 
curvature, like $A$ has). The embedding $s$ is a section of $\End(L, E)$; 
$A'$ and
$A$ induce on this bundle a connection $B$, and the condition 
that the subbundle be
holomorphic can be explicitly expressed as $\bar D_B s=0$ (here, as above,
$D$ 
denotes the holomorphic part of the covariant derivative). Finally, let us put
a hermitian metric $(\cdot,\cdot)$ on the bundle $\End(L,E)$.
We can now consider the equalities
\[
0= \int \partial (D_B s, s)= \int (\bar D_B D_B s, s) -
\int (D_B s, D_B s).
\]
Since, in the form notation we have been using so far, 
$ D_B \bar D_B + \bar D_B D_B = [D_B ,\bar D_B ]
= F_B^{(1,1)}= F_B$, and $\bar D_B s = 0 $, we have 
\[
\int (D_B s, D_B s) = \int (F_B s, s) = 
(2\pi i)\frac{\int (s, s)\omega }{Vol} (\mu(E)- \mu(L));
\]
but, since the (imaginary part of the) 
lhs is non negative, we have that if on $E$ there is a 
holomorphic connection that satisfies $F = c \omega$, $E$ satisfies the 
following property, called $\mu$-semistability: {\sl 
for any holomorphic subbundle $L \hookrightarrow E$, one has 
$\mu(E)\geq \mu(L)$}. Sufficiency of this property, in this and in the other 
cases discussed in the mathematical literature, can also be shown using 
moment maps in the infinite-dimensional space of connections. A
similar, slightly stronger notion is $\mu$-stability, for which we
simply substitute $\geq$ with $>$.

This simple equation is the lowest form of life in the zoo
of equations we have been considering so far. We have now to consider several
generalizations of this. First of all, we can go up in the dimension. The 
equation becomes now (\ref{hym}), and the analogue of what we have seen above
in 2 dimensions works perfectly; $\mu$ is now $deg / rk$, where $deg = c_1 
\cdot [\omega]^{n-1}$ depends now on the K{\"a}hler class $[\omega]$.


Since in general the definition of $\mu$-(semi)stability involves 
checks of the inequality for all coherent sub{\it sheaves} $\EE'$ of 
rank  $0<rk(\EE')<rk(E)$, the argument above may seem to have a caveat.
Fortunately it is sufficient to carry out the check for only so-called 
reflexive sheaves  \cite{uhlyau}, and the only such sheaves of rank one 
are  line bundles. More generally, it is shown in \cite{luebke} that HYM 
implies stability even in the stronger sense (that is, including 
subsheaves). 
One however needs stability with subsheaves in order to show that 
stability implies the existence of a HYM \cite{uhlyau}. 
Low dimensions (two and four) are exception \cite{donkro}, and 
the definition of the stability used above is adequate. 
Finally, there are other
more refined notions of stability; we have in mind 
{\it Gieseker} stability in particular. It is related  
\cite{leung}  to an asymptotically
large-volume analysis of deformed equation (\ref{mmms}). 
One of the differences between Gieseker and $\mu$ stability is that now
subsheaves are not even required to have smaller rank;
see for example \cite{huylehn} for a more detailed discussion.
Similar considerations to the ones in this paragraph should be taken
in account whenever we speak of subbundles.

\vspace{.3cm}
\noindent
{\it Deformations}
\vspace{.2cm}

Now we turn to the deformations (\ref{mmms}). These are expected to 
be related, 
modulo some corrections, with $\Pi$-stability 
\cite{Douglas:2000ah,Douglas:2000gi}. 
The latter is a stability 
for branes very similar in spirit to the stabilities considered by 
mathematicians (the similarity can indeed be summarized using
the powerful mathematical language of categories): for each subbrane
$B'$\footnote{In general, $\Pi$-stability is actually defined in terms of
distinguished triangles in the derived category \cite{Douglas:2000gi}; 
it is however tempting to examine the possibility that also in other
areas of the moduli space the analysis can be reduced to one which
only involves a simpler {\it abelian category}, that is, one in which
it still makes sense to talk about subobjects. These subobjects are
the things that we will call subbranes in this paper.}
of a given brane $B$, the relation $\phi(B)\geq
\phi(B')$ should be satisfied, where $\phi$ is, modulo some subtleties, 
the argument of the central charge of the brane.
This stability interpolates between different known stabilities in 
various limiting points of the moduli space; in particular, in the limit in 
which branes can be considered as geometric objects 
(holomorphic cycles with bundles on them), the inequality with $\phi$ 
reduces to the inequality with $\mu$, and we recover the usual stability for 
bundles. This is clearly similar to the 
way in which (\ref{mmms}) reduces to (\ref{hym}) in the limit of large $\omega$
(or small $F$). To be more precise, let us underline once again the reason for
which $\mu$ is appearing in the usual stability. The reason is that $\mu$ is 
proportional to the constant $c$ appearing in HYM (\ref{hym}). 
Now, we have a constant in  (\ref{mmms}) as well: it is $\theta$. 
As in the HYM  case, the way to uncover the relation of this with the 
topological constants 
is to take the trace of the equation and integrate. Here we get
\bea
&&\nonumber 
\int \mat{Im} (e^{i \theta} e^{F + \omega})_{\mat{top}} = \\
&&\nonumber 
 \sin(\theta) \int ( \omega^n + \omega^{n-2}\mat{Tr} F^2 + \ldots) +
\cos(\theta)  \int (\omega^{n-1}\mat{Tr}(-iF)+\omega^{n-3}\mat{Tr}(-iF^3) 
+\ldots)=\\ 
&&\nonumber 
 \sin(\theta) \mat{Re} \int ( \omega^n + \omega^{n-1}\mat{Tr} F + \ldots ) +
\cos(\theta)  \mat{Im} \int ( \omega^n + \omega^{n-1}\mat{Tr} F + \ldots)= \\
&&\nonumber 
\mat{Im} \left(e^{i \theta} \int e^{F + \omega} \right) 
\eea

The equation (\ref{mmms}) sets this to zero; thus we find that 
$\theta= \mat{arg}(\int e^{F + \omega} ) (+ \pi k i)$. Since in the
trivial geometries, this expression is the same as the central charge, this
suggests that the corrections \cite{dougag}
to the equation in a nontrivial geometrical setting amount to
introduction of an additional factor 
$e^{- F_a/2}\sqrt{\hat A(R)/\hat A(F_a)}$, where $a$ is the connection on the
normal bundle, $F_a$ its curvature and 
$R$ the Ricci curvature on the brane. In this way we get 
$\theta = \mat{arg}\left(\int e^{F + \omega} 
e^{- F_a/2}\sqrt{\hat A(R)/\hat A(F_a)} \right) = \mat{arg}(Z) = 
\phi$. Unfortunately, it is now harder to reproduce the proof we saw for HYM
due to the nonlinearity of equation (\ref{mmms}). 
One of the central points of that proof was the appearance of $F$ from the 
anticommutator of two covariant derivatives; the analogue of this is not 
totally clear.  The lack of supersymmetric completion for CS terms
involving gravitational corrections is another obstacle in this
direction. One could try a different approach by attempting to
generalize the usual moment map method (see e.g. \cite{thomas}) but this
is beyond the scope of our work.

\vspace{.3cm}
\noindent
{\it Transverse scalars}
\vspace{.2cm}

Let us come now to the main theme of this paper, the introduction of the
transverse scalars.  Foreseeing problems due to non-linearity of
the equations, one can first introduce scalars in the  ``linearized''
case, the HYM, and perform the analysis there. This should serve as a base
for extending the modifications to the non-linear equations.
To be specific, let us concentrate on the Hitchin equations in four
dimensions (\ref{Hcov4}) and find a stability condition for this
equation.
We will denote by $\langle\, ,\rangle$ the inner product $\int vol \,(\, ,)$,
with $vol$ the volume form, and make use of the Weitzenb{\"o}ck equalities
\cite{donkro}
\[ 
\frac12D_A^\dagger D_A = \partial_A^\dagger \partial_A -i\,\omega\cdot F_A\ .
\] 
As in the easier case  of two dimensions and no scalars, discussed in the
beginning of this subsection, we will choose for simplicity a rank 2
bundle $E$, and consider a sub-line-bundle $L$ on which,
without loss of generality, we can put
a connection $a$ with constant curvature. Applying a Hodge star (and changing
$c$)
we rewrite (\ref{Hcov4}) in the form
\be 
i\,\omega\cdot F + \sum_{a=1}^3\, [X_a,X_a^\dagger]=c\ .
\label{again}
\ee
On the bundle $\Hom(L,E)$ we can now consider again the connection $B$ 
induced by the 
connections $a$ and $A$ on $L$ and $E$. We can apply this covariant derivative
again to the holomorphic section $s$ of $\Hom(L,E)$ expressing the subbundle 
relation (the embedding):
\[ \frac12\langle D_B^\dagger D_B\, s, s\rangle =  \langle \partial_B^\dagger
\partial_B\,s,s\rangle - i\,\langle\omega\cdot F_B \,s, s \rangle\ .\]
Using adjunction (that is, integrating by parts), 
holomorphicity of $s$ and (\ref{again}) we have
\be
 2\pi \frac{\langle s, s\rangle}{Vol} (\mu(E)- \mu(L)) =
\langle D_B s, D_B s\rangle + \sum \langle [X_a,X_a^\dagger]s, s\rangle\ .
\label{proof}
\ee
Suppose  now that $X s = \lambda s$; this equation can be seen in a
sense as looking for an ``eigenvalue'' $\lambda$ which is actually a section
of $N(B,M)$. With a little abuse of language we will be calling  $s$ 
an ``eigenvector''. Then we have for the last term
\[ \langle [X, X^\dagger]\,s, s\rangle= 
\langle X^\dagger\, s, X^\dagger\, s\rangle
- |\lambda|^2\langle s, s\rangle =\langle(X^\dagger - \bar\lambda)s,
(X^\dagger - \bar\lambda) s\rangle \ .\]
Since this  is non-negative, from (\ref{proof}) we get the condition we 
wanted: {\sl for any subbundle $L\buildrel s\over\hookrightarrow E$ which
is $X$-invariant} (namely, the embedding $s$ is eigenvector of all the
$X_a$'s) {\sl we have the relation $\mu(E)\geq\mu(L)$}. This is the  
four-dimensional analogue of Hitchin stability. We have here only shown
that it is necessary for solution of (\ref{Hcov4}) to exist; a proof of
sufficiency is more difficult, but essentially standard along the lines
of other linear examples, using orbits of the complexified gauge group and
analytic arguments, as for example in \cite{uhlyau,Hitchin:1987vp}.

A similar case to this one, though with a different ``twisting'', has
been extensively studied in the mathematical literature under the name 
of Higgs bundles \cite{higgs}.
\vspace{0.2cm}

Moving now finally to our equations (\ref{nice}), it is very natural to propose
a Hitchin-like modification of stability, in a more general abelian
category (see previous footnote: what we need here is a category in
which it makes sense to speak about exact sequences): for each 
$X$-invariant holomorphic subbrane $B'$ of a given brane $B$, one has
$\phi(B)\geq\phi(B')$. The only step left is to specify what 
``$X$-invariant'' is supposed to mean, since in general (away from
geometrical limits in the 
moduli space) branes are not bundles on cycles (that is, we are in a
different abelian category). For this, we notice that 
one can reformulate this notion in a completely abstract way by completing
the embedding $B' \hookrightarrow B$ to the
exact sequence 
\[  0 \to B' \buildrel i \over\longrightarrow B 
\buildrel\pi\over\longrightarrow B'' \to 0.  \]

Using the maps $i$ and $\pi$, one can see that being $X$-invariant means that 
$\pi \circ X \circ i= 0 $ as an element of $\Hom(B, B'')$, where 
$\circ$ is composition.  We emphasize again that the status of the 
abelian category in question is rather uncertain; all we know is that its 
objects should involve in some sense the
cycles and their embeddings in $M$, or equivalently the normal
bundles.  The morphisms in this category should also contain
these data, in order to match  the known large-volume limit, 
in which the category is the one of sheaves and the morphism $X$ is a
section of $End(E)\otimes N(B,M)$. 
Though this proposal is the natural melting
of $\Pi$-stability and of Hitchin's one, let us stress that again, as
for the proposed connection between $\Pi$ stability and (\ref{mmms}), 
a complete proof is very difficult to find due to the non-linearity of the
equations.

We would like to
underline that these modifications {\`a} la Hitchin to the usual
stabilities is substantial. Once one fixes a cycle $B$, a bundle 
$E$ on it and a endomorphism $X$, it can happen in fact that even 
if $E$ was unstable with respect to the usual definition, it is stable with
respect to 
the modified one. Let us analyze for example $\mu$-stability. Suppose 
that there exists only one holomorphic subbundle $E'$ which destabilizes $E$,
$\mu(E')>\mu(E)$. Then, if this subbundle is not $X$-invariant, $E$ is still
$\mu$-Hitchin-stable. (Or more accurately, the couple $(E,X)$ is
$\mu$-Hitchin-stable.)

\vspace{.3cm}
\noindent
{\it Quivers and $\theta$-stability}
\vspace{.2cm}

A logical consequence of this is also that the $\theta$-stability that one 
finds going to the Gepner point \cite{Douglas:2000ah, king} should be modified
in a similar fashion. Let us give a quick look at this here. 
First of all, recall that, near the Gepner point of the moduli space, branes
are described to some extent by the same approach describing branes on an 
orbifold singularity \cite{Diaconescu:2000ec}. Thus one has a supersymmetric
gauge theory whose gauge content is 
summarized by a quiver; here, however, we will not need this explicitly, and
keep all of the chiral multiplets in a set of total chiral fields $\Phi_i$
whose blocks are the matrices which represent the quiver.
Then, we can write the D-term and F-term equations which describe the 
moduli space of the theory as 
\be
\sum_a [ \Phi_a, \Phi_a^\dagger ] = \Theta \mat{Id}; \qquad 
\frac{\partial {\cal W}}{\partial \Phi^a} = 0 \label{D-F}
\ee
As in the geometrical limit, the F-term equation (the
$(2,0)$ part) is holomorphic  and does not get modified by K{\"a}hler
moduli. The equation that leads to a  stability condition is again the
single real equation for D-flatness, the first one in (\ref{D-F}).
The block-diagonal matrix $\Theta$ contains FI terms. 

The condition for existence of solutions for these equations, again
implies a stability
condition. Indeed, consider a subrepresentation of the quiver. From the point
of view of the total fields $\Phi_a$ (of rank, say, $k$), 
this means there is an injection $i$
such that smaller matrices $\Phi_a'$ of rank $k' < k$, satisfying 
$\Phi_a \circ i = i \circ \Phi_a'$ can be found (we will omit $\circ$ from now
on). It is useful to introduce as well a matrix $i^\dagger$ such that
$i^\dagger i = \mat{Id}_{k'}$; then $ii^\dagger$ is a rank $k'$ projector $p$,
and we can rewrite $\Phi_a' = i^\dagger \Phi_a i$.
Having introduced such a notation, let us see what happens each time we have
a subrepresentation $i$. Then we start from a quantity manifestly non negative
and expand it:
\bea
0&\leq&\sum_a \mat{tr}
\left( (p \Phi_a (1 - p) )(p \Phi_a (1 - p) )^\dagger \right)=
\label{ganzo}\\
\nonumber&&\sum_a\mat{tr} (p \Phi_a (1 - p) \Phi_a^\dagger)= 
\sum_a\left(
\mat{tr} (p \Phi_a \Phi_a^\dagger)-\mat{tr}(p \Phi_a p \Phi_a^\dagger) \right);
\eea
but then, being
\bea
\nonumber \mat{tr} (p \Phi_a p \Phi_a^\dagger)&=& 
\mat{tr} (ii^\dagger \Phi_a ii^\dagger \Phi_a^\dagger)=
\mat{tr_{k'}} (i^\dagger \Phi_a i\, i^\dagger \Phi_a^\dagger i)=
\mat{tr_{k'}}(\Phi_a' \Phi_a'^\dagger) = \\
&&\mat{tr_{k'}}(\Phi_a'\Phi_a'^\dagger i^\dagger i) = 
\mat{tr}(\Phi_a p\Phi_a^\dagger),
\label{steps}
\eea
we can reexpress (\ref{ganzo}) as 
\[
\sum_a\mat{tr} (p [\Phi_a,\Phi_a^\dagger]) = \mat{tr} (p \Theta).
\]

So, each time we have a subrepresentation of the quiver, the relation 
$\mat{tr} (p \Theta) \geq 0$ should hold. This is called $\theta$-stability.
This direct and explicit proof of necessity of stability for solving the
D-term equations is exactly along the lines of the one we gave at the
beginning of this long section. Now, it is not evident that the modification
{\`a} la Hitchin of the stability that we have introduced will survive till this
point of the moduli space; after all it is not clear to what the
endomorphism which was called $X$ in
the geometrical limit will correspond in the quiver limit.
In particular it could correspond to the zero endomorphism, thus giving no 
modification at all. But, at least mathematically, we can give an example of
an equation whose solutions would imply a $\theta$-Hitchin-stability.

Thanks to this analysis, this is now almost trivial.  First, one
must set the condition that the quiver representation is
$X$-invariant --  the equation $[X, \Phi_a]=0$.  Then, one finds the
appropriate modification to the D-term equation in the form
\be
\sum_a [ \Phi_a + \alpha_a X , \Phi_a^\dagger + \alpha_a^* X^\dagger] 
= \Theta \mat{Id} \ .
\ee
Indeed, for a subrepresentation to be
$X$-invariant, the injection $i$ should satisfy the additional condition 
(similar to the above with $\Phi,\Phi'$) that a smaller matrix $X'$ should 
exist, such that $X i = i X'$. If a subrepresentation satisfies this, there is
no problem in carrying out the steps (\ref{steps}) in exactly the same way,
with $ \Phi \to \Phi + \alpha X$; otherwise this is not possible. Thus,
the result is that we have the inequality $\mat{tr} (p \Theta) \geq 0$  only
for $X$-invariant representations. 

Thus we obtain a modification to $\theta$-stability very similar to ones
in geometric phase, which again may in particular rehabilitate representations
previously discarded as unstable. One can view this as the 
survival of Hitchin-like modification all over the moduli space.
We will see, however, that from the point of view of the moduli space of 
all the BPS states, this modification is not so dramatic as one could think, 
thanks to the fact that branes can be lifted to coverings, as we will now 
explain.

\vspace{.3cm}
\noindent
{\it Coverings}
\vspace{.2cm}
 
The starting point 
for this technique (which has been used already for instance in 
\cite{mst,Bonelli:2000nz})
is the standard interpretation for non-abelian
transverse 
scalars. Their virtue is that they allow to describe, in different vacua, 
configurations with several branes wrapped on the same locus 
or the same number of branes, but each wrapped in a different
locus. For instance, 
the vacuum $X=0$ represents $N$ (the rank of the matrix) branes wrapped on 
the submanifold $B$ on which the non-abelian brane theory is defined;
whereas 
the vacuum $X = \mat{diag}(x^{(1)}, \ldots, x^{(N)})$ with 
$x^{(1)}\neq \ldots \neq x^{(N)}(\neq 0)$ describes again $N$ branes, 
but displaced from one another and from the ``initial locus'' $B$. 
Usually the configurations for the transverse scalars are taken
constant, because one wants to consider the vacua. 
Here we will, however, consider general solutions to our equations. 

This time, the most relevant feature we will need, that we have not exploited 
so far, is the fact that the $(2,0)$ part of the equations does not get 
deformed. This part is (compare with (\ref{sdH4})) 
\be 
F^{(2,0)}=0, \qquad [X_a, X_b]=0, \qquad D X_a=0 \label{f20}
\ee
We will for most of the time analyze the case with only one transverse 
scalar $X$, and come later to more general cases.
Let us begin from the consequences of last equation in (\ref{f20}), 
which is now $D X=0$; and
let us first pretend we are in flat space.
We have recalled the standard interpretation
that the eigenvalues of $X$ give the classical position of the branes. These
eigenvalues
are given by the roots of the characteristic polynomial $p_X(x) = \det(X - x)$;
last equation in (\ref{f20}) implies $\partial p_X(x)= 0$. 
Then this characteristic polynomial
can be viewed as an equation $p(z_1, \ldots, z_n, x) = 0$ in 
$B\times \cc = \cc^n \times\cc$,
and thus cuts out a complex $n$-dimensional locus. This is the same 
dimensionality of the base, as it should, and in fact this locus is a $N$-fold
covering of $B$: over a generic point of $B$, 
there are $N$ counterimages, and so locally this looks 
the same as the usual picture of several branes, wrapped on different loci,
described by the eigenvalues of $X$ (our previous example 
$X = \mat{diag}(x^{(1)}, \ldots, x^{(N)})$). But this time, apart from the 
most trivial case in which $p$ is constant in the $z_i$'s, the covering will 
be branched; that is, 
the eigenvalues will come to coincide somewhere, and thus the 
resulting covering brane will be {\sl one}. This is the so-called spectral
manifold \cite{Hitchin2,Hitchin3}.

Summarizing: in generic cases, the eigenvalues of $X$ describe a single brane
$\tilde B$
which is geometrically a branched covering of the base $B$. 
We have described this
for the flat space case for simplicity; but this so-called spectral manifold
can
be defined also in the most general case, in which the base manifold $B$ 
on which the non-abelian
brane theory is defined is an arbitrary manifold, and $X$ is a section of 
$N(B,M)\otimes\End(E)$. The equation $p_X(x)=0$ becomes now an equation in the 
total space of the line bundle $N(B,M)$ (or $K_B$, which is the same, if the
ambient manifold $M$ is a Calabi-Yau). Now $DX =0$ means, as we have
underlined before the beginning of subsection \ref{twi}, $dX + [A, X] + a X
=0$, where $A$ is the part of the connection in $\End(E)$ and $a$ is a
connection on the bundle $N(B,M)$. Thus we have now $(\partial + a) p_X(x)= 0$;
this can be considered as giving a holomorphic structure to the submanifold
 $p_X(x)=0$ in the total space of the bundle $N(B,M)$.
With these modifications, 
the geometrical interpretation is intuitively the same.

Actually, what we want is not quite a submanifold defined on the total space
of the normal bundle $N(B,M)$; 
this is only a local (around the initial brane $B$) 
description. Then we can make $\tilde B$ a submanifold of the ambient
manifold via the map $N(B,M) \to M$, as it is done in K-theory to find Gysin
map from the Thom isomorphism. 

Finally, as noted many times to specify a submanifold is not enough for
giving a brane, we have to  provide the bundle on it as well. Since one
expects the covering brane 
$\tilde B$ to
be a single object, pulling back the non-abelian gauge bundle from the
base $B$
would not be a good choice: we want a line bundle. The right construction is 
remarkably simple \cite{Hitchin2,Hitchin3}. 
The line bundle on the covering brane $\tilde B$ can be constructed recalling
the very definition of $\tilde B$: over each point of the base $B$, we consider
the $N$ eigenvalues of $X$ as values of the extra transverse coordinate.
The line bundle is defined by considering, 
on each point of the covering (which is an eigenvalue), the corresponding {\sl 
eigenspace}. This line bundle $L$ is natural in the sense that its push-forward
from $\tilde B$ 
to the base $B$ 
(this operation can be defined by resorting to sheaves) is indeed
the gauge bundle on the base: $\pi_* L = E$, where $\pi: \tilde B \to B$ is 
the covering map.

\vspace{.3cm}
\noindent
{\it Coverings, stability and the BPS states}
\vspace{.2cm}

We have now a non-abelian brane theory defined on a submanifold $B$ 
of a given ambient
manifold $M$. We have argued that stability gets modified {\`a} la Hitchin, with 
the condition on subbundles (or subbranes) replaced by the same condition 
on $X$-invariant subbundles. 

The first case is the one in which $X=0$. In this situation the branes are
all wrapped on the base manifold $B$. 
But in this case, the modification we have
proposed does not affect anything, because being $X$-invariant is an
automatically satisfied condition.

If, on the other hand, $X$ is a non trivial configuration, then we have
seen that this describes branes on a covering $\tilde B$ of the 
initial brane $B$. In
particular, we have seen how we can lift the gauge bundle from $B$ to $\tilde 
B$. This is important for the following reason. If one considers BPS states on
a fixed base $B$, there are substantial contributions coming from the presence
of the scalars $X$, as we have seen. But, if one considers the whole moduli
space of BPS states, namely the union of the former moduli space for all the
base branes $B$, the covering technique actually shows that most of the new
BPS states one obtains for one base brane $B$ are actually copies of other
BPS states pertaining to another brane $\tilde B$\footnote{We may
  present another somewhat indirect argument in favor of this
  assertion by recalling the definition of $\phi(E)$.
As it was argued in \cite{Hassan:2000zk}  taking into
account the non-abelian dynamics of D-branes amounts to replacing
the RR coupling $C \wedge Y$ by Clifford multiplication of $Y$ by RR
fields while leaving intact the form of the D-brane charge $Y$, and thus the
element in $K(M)$. We should note however that transforming CS
data into equations of motion is not straightforward in view of
difference in ``natural" conventions for the kinetic and CS parts. We
will meet another similar clash of conventions in section \ref{dte}.}.

The discussion we had so far has some limitations.
First, there are subtleties concerning the covering mechanism.
Indeed, we have supposed so far that the 
eigenvalues are all distinct. If the characteristic polynomial is irreducible
(which means that the covering is connect; this we can assume with no loss of
generality) the only way we have to duplicate eigenvalues is to duplicate 
all of them, taking thus a characteristic polynomial which is a power: 
$p_{X'}(x) = (p_X(x))^k$. In this case, lifting yields a vector bundle 
$\tilde E$ {\sl of rank $\leq k$} over
the covering brane $\tilde B$; a priori this is not guaranteed to be in the
usual moduli space of BPS states pertaining to the cycle $\tilde B$, since
$\tilde E$ could be not stable. 
But the modification of the stability {\`a} la Hitchin that 
we have proposed means now that the 
subbundles $E'$ of the gauge bundle $E$ on the base $B$
should be liftable too. This way,
the stability condition will get translated into a stability for the 
vector bundle $\tilde E$ over the lifted brane $\tilde B$.

More importantly, so far we have analyzed the covering mechanism
in the case in which there is only one transverse scalar. When there are more,
although there is an equation $[X_a,X_b]=0$ from the undeformed $(2,0)$ part 
of the equations, still the commutators like $[X_a,X_b^\dagger]$ do not
vanish: they appear in the deformed $(1,1)$ equations. This means that in
general we are entering the realm of noncommutative -- perhaps better, fuzzy --
BPS states; we will not analyze this, but for the small note that follows.

\vspace{.3cm}
\noindent
{\it More transverse scalars}
\vspace{.2cm}

The analysis of this case allows us to appreciate the importance of the
part of the connection corresponding to the normal bundle $N(B,M)$. Let 
us first of all tackle the flat space situation in which the brane is at
least 2 dimensional but there are more than one complex scalar.

In this case our equations have a strange feature. 
If we consider more than one transverse scalars, again because of the
equation $D X_a = 0$, 
all of them will have an antiholomorphic characteristic
polynomial, $\partial \det (X_a - x_a)= 0$. But the second equation of 
(\ref{f20}) implies, in the generic case in which the eigenvalues are
not coincident, that the matrices are simultaneously diagonalizable. This in
turn implies that the monodromies of the eigenvalues are the same, and thus 
that the characteristic polynomial are the same \cite{Bonelli:1999it}. 
In this situation the $X_a$
turn out to be multiples of each other. In general, we can relax 
the assumption that the eigenvalues are distinct, and find situations in which
the matrices are not multiples of each other; so it still makes sense to keep
(\ref{nice}) in its form with several scalars. But the characteristic 
polynomials remain the same, and so in a way for the covering considerations 
we have done before only one of the scalars is sufficient.

Let us now see what differences come in when we are in a geometrically less
trivial situations; let $B$
be a brane wrapped on a 2-cycle (or higher) on the ambient manifold $M$.
(In the main example we have had in mind, that of $M$ being a \CY threefold,
the only such case is the one in which $B$ is a 2-cycle.) Then, as we have
seen in the case with one scalar only, 
the equation
$DX=0$ does not imply any longer that $\partial \det (X_a - x_a)= 0$, but
that $\partial_{N(B,M)} \det (X_a - x_a)= 0$, where 
$\partial_{N(B,M)}$ is a connection on $N(B,M)$. This prevents the arguments
we expounded in the paragraph above to be applied in this more general
case. Finally, we can 
consider instead the case in which $B$ is a 0-cycle. Then the above
considerations do not apply in flat space. 
The equation $D X_a = 0$ becomes now one of the
already present equations $[X_a, X_b]=0$, and so all of its consequences that
we
have explored in the preceding paragraph are not there any longer. In this
case we are thus completely in the realm of fuzzy solutions; but we will not
explore this here.

\newsection{Deformed tachyon equations}
\label{dte}
In this section, we use again dimensional reduction, this time in a different
way: we will obtain equations that we will argue to be relevant for the 
tachyon of the D$p-\overline{\mat{D}p}$ system. 

Let us consider a pair of gauge bundles, $E_1$ and $E_2$, describing the
gauge theory on the brane and on the antibrane, and a morphism of bundles
$T$ connecting them, that can be thought of as a section of $\End(E_1, E_2)$.
In \cite{Oz:2001bs}, a  set of equations has been described  
that implies the equations of motion
of this D$p-\overline{\mat{D}p}$ system, much in the same way in which the 
instanton equations imply the equations of motion for Yang-Mills, and more
generally in the same way in which BPS conditions imply equations of motion
for a supersymmetric action. These equations read
\be
F_1 \cdot \omega -i T T^\dagger  = \lambda_1 \mat{Id}_{N_1}, \qquad  
F_2 \cdot \omega +i T^\dagger T  = \lambda_2 \mat{Id}_{N_2}, \qquad  
\partial T + A^{(1,0)}_1 T - T A_2^{(1,0)}  = 0, \label{tac}
\ee
where $N_i = rk(E_i)$; along with the usual $F^{(2,0)}$. 
As for the HYM equations and their deformation we have been studying so far,
solution to them is equivalent to a stability condition on the triple 
$(E_1, E_2, T)$. 

Now, the point interesting for us is that these equations come naturally 
again from dimensional reduction of HYM, although in a more formal way. 
Namely, we reduce from complex dimension $n+1$ to $n$, in such a way that only
one complex scalar will appear; this we will call $W$, as it is not 
one of the scalars parameterizing transverse fluctuations (for instance, the 
procedure we are following will give us the tachyon equations also in the 
maximal dimension $n=5$, that is for the D9-$\overline{\mat{D}9}$ system, in
which case there are no transverse fluctuations at all).
Another modification to the reduction
we have done before is that we have now to take a slightly more general 
non-abelianization of them, considering a direct sum bundle; thus, in the
rhs
of (\ref{hym}), we can be more general and write, instead of $\lambda \,Id$, 
$ diag(\lambda_1 Id_{N_1}, \lambda_2Id_{N_2})$ ($diag$ is here in a block
sense). 
With this specifications, let us now do dimensional reduction of HYM
(\ref{hym}) taking the
particular choice\footnote{This choice can be better 
motivated \cite{bradlow} if one takes the 
fibre to be $\pp^1$, and chooses then the bundle on the whole fibration to 
be $p_1^* E_1 \oplus  p_1^* E_1 \otimes p_2^*\oo(2)$. Here we are doing a 
formal reduction as in sections \ref{pre} and \ref{eqns}.}
\be 
A = \left(\ba {cc} A_1 & 0 \\ 0 & A_2 \ea \right),\qquad 
W= \left(\ba {cc} 0 & T \\ 0 & 0 \ea \right) \label{choice}
\ee
(remember that $W$ has the formal role that $X$ had in previous sections,
appearing in Hitchin equations in its stead).
In this way one obtains precisely the tachyon equations (\ref{tac}): first two
come from the $(1,1)$ part, the third comes from the $(2,0)$ part. 

We have obtained these tachyon equations from reduction of HYM; but we have
seen that the latter get deformed by stringy corrections to
(\ref{mmms}). 
Thus we expect that, if we now reduce (\ref{mmms}) in the same way, 
we will obtain
the right deformation of the tachyon equations. This is by now an easy task:
first of all, as usual, the $(2,0)$ part does not change, so we will keep on 
getting $DT \equiv \partial T + A_1^{(1,0)} T - T A_2^{(1,0)} = 0$. 
Then, let us start from our expression 
(\ref{nice}) and first of all specialize it to the case in which there is only
one complex scalar. We get 
\be
\mat{Sym} \{ \ (e^{F + \omega \mat{Id}})_{\mat{top}}(1 -i [W , W^\dagger ]) + 
\frac i2\left([D, W^\dagger] [\bar D, W ] - [\bar D, W ] [D, W^\dagger]\right)
(e^{F + \omega \mat{Id}})_{\mat{top}-2} \ \}.
\label{dte1}
\ee
Once more, let us now extend this to the case in which the bundle is 
a direct sum;
in this case this means to substitute $e^{i \theta}$ in (\ref{niceeq}) with 
$\exp\{diag(i \theta_1 Id_{N_1}, i \theta_2 Id_{N_2})\}$. 
Then, putting in this expression the particular choice (\ref{choice}),
one obtains our ``deformed vortex equations''
\bea
&&\nonumber \mat{Im} \left( e^{i \theta_1} \mat{Sym} \{ 
(e^{F_1 + \omega \mat{Id}_{N_1}})_{\mat{top}}(1 -i T T^\dagger ) - \frac i2
\bar D T D T^\dagger (e^{F_1 + \omega \mat{Id}_{N_1}})_{\mat{top}-2} 
\}\right) = 0\\
&& \mat{Im} \left( e^{i \theta_2} \mat{Sym} \{ 
(e^{F_2 + \omega \mat{Id}_{N_2}})_{\mat{top}}(1 +i T^\dagger T ) + \frac i2
D T^\dagger \bar D T (e^{F_2 + \omega \mat{Id}_{N_2}})_{\mat{top}-2} 
\}\right) = 0,
\label{dte2}
\eea
where 
$\bar D T \equiv \bar\partial T + A_1^{(0,1)} T
- T A_2^{(0,1)}$ is the antiholomorphic covariant derivative of the tachyon.
Notice that these equations are less decoupled than usual: not only 
does the tachyon appear in both, but also, in the deformation term with $\bar
D T$, both $A_1$ and $A_2$ appear.

It is natural to speculate that (\ref{dte2}) can be expressed in terms
of superconnections.
Let us try to make this expectation more precise. One would like to exploit 
the considerations made after (\ref{nice}) to write an expression which 
contains a superconnection in the exponent. To do so, the procedure is just
like the one we have done above:  i) replacing all the $Id$ with  
$ diag(Id_{N_1}, Id_{N_2})$, ii) taking
$A$ and $T$ of the particular form (\ref{choice}). But this time, one wants
to start not from the more explicit form of the equation (\ref{dte1}), but 
from one of the more imaginative forms (\ref{nice}) or (\ref{nice2}). 
Explicitly, one gets
\be
\mat{Im}\left( \, \mat{Sym} \{ \exp
\left(\ba {cc}i \theta_1\mat{Id}_{N_1}&0\\0&i\theta_2\mat{Id}_{N_2}\ea\right)
\,[e^\ff e^{\omega + J}]_{\mat{top}} \}\, \right)  = 0,
\label{dtenice}
\ee
where 
\[
\ff \equiv 
\left(\ba {cc} F_1 + i_T \,i_{T^\dagger}& i_T \,\bar D_2 - \bar D_1 \,i_T \\ 
D_2 \,i_{T^\dagger} - i_{T^\dagger}\,D_1& F_2 + i_{T^\dagger}\, i_T \ea\right) 
= [\dd,\bar\dd]; \qquad \dd = 
\left(\ba {cc} D_1 & i_T \\ 0& D_2  \ea\right).  
\]
These equations need of course some comments. First, we remind the 
reader that we are using here a super-Lie bracket instead of the usual one, as
explained after (\ref{nice2}); this explains the sign of $i_{T^\dagger} i_T$
in $\ff$. 

Second, we introduced in (\ref{dtenice})
a symbol $J$ which is a two-form $idwd\bar w$ in the formal transverse space
spanned by $W$, $W^\dagger$. Once again, this is not the physical transverse
space, and this $J$, though it has the same formal role of $\omega_\perp$ in
(\ref{nice}) and similar equations, is a different object, introduced here only
as a convenient device for bookkeeping.
Given that there is only one holomorphic object $T$ here, this choice could 
appear rather baroque; and one could wonder if, instead, it would not have
been better to start from an expression halfway between (\ref{nice}) and
(\ref{dte1}); namely, an expression compact as (\ref{nice}) is, with an
exponential structure, but with $i_W$ replaced in some way by $W$. In fact,
such an equation does not exist. To replace $i_W$ with $W$ (or $i_T$ with $T$
in (\ref{dtenice})) before expanding the exponential would be wrong, as one
can convince oneself by noting the minus sign in the expression 
$[D, W^\dagger] [\bar D, W ] - [\bar D, W ] [D, W^\dagger]$
in (\ref{dte1}). Moreover, the expression one could
obtain this way will necessarily  have a piece proportional to 
$diag(T T^\dagger, - T^\dagger T )$, which comes
from a commutator $[W, W^\dagger]$ (compare with (\ref{tac})). Thus, this 
expression would be in any case different from the superconnection
arising from the CS system of the $D-\bar D$ system. It is hardly
expected that the two superconnections are the same. Since we are using
the equations of motion, there is a contribution from the DBI part.
However, it is interesting to notice that the latter do not spoil
completely the presence of the $\zz_2$-graded structures. 



Finally, let us notice that it could be that (\ref{dte2}), alias
(\ref{dtenice}), is
the right deformation to solve some of the problems of interpretation for
stability of triples raised in \cite{Oz:2001bs}.

\vspace{.3cm}
\noindent
{\it Tachyons and transverse fluctuations}
\vspace{.2cm}

By expanding on the observation in section \ref{eqns} on the similarity of
equations involving the scalars to those appearing in the computation of
Chern character with superconnections, we conclude by a comment on
relation between tachyons $T$ and transverse scalars $X$ on D-branes. So
far this is only a formal correspondence, as $T$ and $X$ are quite
different objects. One may furthermore note that the reduction performed
in this section for the tachyon equations is formal, unlike that for $X$
which was given by T-duality. For instance, there is no reason for which
the tachyon equations we propose should not be valid in the important case
of D9-$\overline{\rm{D9}}$ system (in which case there are no transverse
scalars).

Nevertheless, there is a relation between the two objects, provided by tachyon 
condensation. First of all, let us begin with a couple of considerations on
the general meaning of what we are going to analyze. Consider a 
D$p-\overline{\mat{D}p}$ system on some manifold $M$. This defines, through 
relative K-theory, some element of the $K(M)$. Superconnections were initially
introduced \cite{quillen} 
as a method to compute the Chern character of this element.
If we consider instead a  non-abelian brane theory defined on some
submanifold $B$, by the covering 
mechanism this describes a brane wrapped on some other submanifold $\tilde B$
of $M$. This is again an element of $K(M)$. Tachyon condensation
connects these two ways of obtaining a K-theory element; the
result, as we will see shortly, is that $T$ is the characteristic
polynomial of $X$.

To be more specific, consider a D9-$\overline{\mat{D}9}$ 
system on $\cc\times M$, where 
$M$ is now a 8 dimensional manifold. Let us first consider the case in
which 
the bundles on the brane and the antibrane are chosen in such a way that 
the tachyon is $T=x$, where $x$ is the coordinate on $\cc$. 
Thus the system condenses to a D7 wrapped on $M$, described by the locus
$x=0$. 
What we want to emphasize is that both $T$ for the D9-$\overline{\mat{D}9}$ 
and $X$ for the D7, the vanishing locus is the same, although from a
different
perspective: in the first case the tachyon is a 10 dimensional field whose
zeroes indicate where the resulting lower-dimensional 
brane will be; in the second case $X$ 
it is a 8 dimensional, whose zeroes indicate in which position of the 
transverse direction that it parameterizes the vacuum is located. Let us go 
ahead with this and consider now a less trivial case: let the tachyon
be $T(x,z_i)$, 
where $z_i$ are coordinates on $M$;
let us suppose it is holomorphic, and moreover that is a polynomial in $x$
of degree $N$. Now, the D7 resulting after condensation can be obtained as
a classical configuration in the non-abelian brane theory with base
$B=M$.
This is accomplished by the covering mechanism we have described above: it
is sufficient to take a configuration $X$ with characteristic polynomial
$p_X(x, z_i)= T(x, z_i)$. Thus, more generally we can say that $T$ is the 
characteristic polynomial of $X$.

Let us give a geometrical twist to this. We can describe this $\cc
\times M$ as a trivial line bundle over $M$; more generally we can consider 
a different line bundle $L$ on $M$, and take the 10 dimensional space as the 
total space $Q$ of this line bundle. 
In this situation the locus is again described
by the zeroes of the tachyon $T(x,z_i)$; but the tachyon itself now is a 
section of the $N$-th power of the bundle which is the pull-back of the 
bundle $L$ to its own total space $Q$. This bundle has an obvious 
so-called tautological
section, and $x$ is to be understood now as this section.

Finally, it is now an easy generalization to consider the case in which the
result of the condensation has codimension higher than 2: 
locally around the lower-dimensional brane, the tachyon will be in that 
case
\be
T= \sigma_a \det( X_a - x_a ),
\label{char}
\ee
where $\gamma_a =\left(\ba {cc} 0& \sigma_a \\0&0\ea\right)$ is the $\gamma$ 
matrix relative to the holomorphic coordinate $x^a$.
This formula makes more explicit in general the relationship between tachyon 
field and transverse scalars we have been using implicitly in deriving 
(\ref{dte2}) and (\ref{dtenice}).

\vspace{.4cm}
\noindent
{\bf Acknowledgements}
\vspace{.2cm}

We would like to thank  G.~Bonelli, L.~Bonora, U.~Bruzzo,
L.~Cornalba and P.~Kaste for useful
discussions, and M.~Douglas, N.~C.~Leung, Y.~Oz, T.~Pantev and R.~Thomas 
for useful e-mail correspondence.
The work of R.~M.~is supported in part by EEC contract
HPRN-CT-2000-00122. A.~T.~would like to thank Ecole Polytechnique for 
hospitality during the initial stage of this work.

\end{document}